\documentclass[aps,amsfonts,nofootinbib,twocolumn]{revtex4}
\usepackage{graphicx}
\usepackage{amsmath}
\def\be{\begin{equation}}
\def\ee{\end{equation}}
\def\bea{\begin{eqnarray}}
\def\eea{\end{eqnarray}}
\def\rx {\rho_{\text{eff}}}
\def\orc{\Omega_{r_c}}
\def\om{\Omega_{\text{m}}}
\def\ol{\Omega_\Lambda}

\begin{document}

\title{Observational constraints on phantom-like braneworld cosmologies}

\author{Ruth Lazkoz$^1$, Roy Maartens$^2$, Elisabetta Majerotto$^2$}

\affiliation{\vspace*{0.2cm} $^1$Fisika Teorikoa, Euskal Herriko
Unibertsitatea, 48080~Bilbao, Spain\\
$^2$Institute of Cosmology \& Gravitation, University of
Portsmouth, Portsmouth~PO1~2EG, UK \vspace*{0.2cm}}

\date{\today}

\begin{abstract}

We investigate a simple braneworld model in which the universe
contains only cold dark matter and a cosmological constant, but
the effective dark energy is phantom-like because of
extra-dimensional gravity effects. Modified gravity screens the
cosmological constant $\Lambda$, allowing for a larger $\Lambda$.
In practice, observations do not favour any significant screening.
We use supernova data, the cosmic microwave background shift
parameter, and the baryon oscillation peak in the galaxy
distribution to constrain the model. We find the mean value of
$\om$ with 68\% confidence limits, and an upper limit on $\ol$ at
the 68\% confidence level. The best-fit model is very close to a
standard LCDM model, but the LCDM model provides a better fit
since it has one less parameter.

\end{abstract}

\maketitle

\section{Introduction}

There is increasingly strong evidence for the acceleration of the
late-time universe, from observations of supernovae, cosmic
microwave background anisotropies and the large-scale structure.
This late-time acceleration poses a major theoretical challenge in
cosmology~\cite{Padmanabhan:2006ag}. Within the framework of
general relativity, the acceleration originates from a dark energy
field with effectively negative pressure: $w \equiv
p/\rho<-{1\over3}$. (It is also conceivable, though it remains
controversial, that acceleration may be due to nonlinear
backreaction or averaging effects~\cite{Kolb:2005me}.) The
simplest model, with cosmological constant or vacuum energy
($w=-1$) as dark energy, i.e., the LCDM model, provides a very
good fit to the data~\cite{Spergel:2006hy}, but the unnaturally
small and fine-tuned value of $\Lambda$ cannot be explained within
current particle physics. Quintessence models, i.e. scalar fields
with $w>-1$, allow for richer dynamical behaviour, but do not
improve the fit to the data, and also do not lessen the severe
theoretical problems faced by LCDM. ``Phantom" models, with
$w<-1$, violate the null energy condition $\rho+p\geq0$, and as a
result, the total phantom energy density grows with expansion.
Phantom scalar fields share the theoretical problems of
quintessence, but in addition, they have negative kinetic energy
and lead to an unstable quantum vacuum~\cite{Carroll:2003st}.

Current observations are compatible with $w<-1$. The WMAP 3-year
data, in combination with large-scale structure and SN data,
allows for $w<-1$ in a general relativistic model: for constant
$w$~\cite{Spergel:2006hy},
 \be
w=-1.06^{+.13}_{-.08}\,.
 \ee
The simplest way to produce $w<-1$ is a phantom scalar field in
general relativity. But the price is instability. Thus it is
interesting to investigate other models in which $w<-1$, but
without negative kinetic energy or associated instabilities. This
can happen if some other effect mimics phantom dynamics, i.e.,
leads to an effective $w_{\text{eff}}$ such that
$w_{\text{eff}}<-1$, but without the presence of a pathological
phantom field.

For example, Boisseau et al.~\cite{Boisseau:2000pr} showed that
this can happen in scalar-tensor theories. A number of other
possible mechanisms for mimicking phantom behaviour has recently
been investigated~\cite{Liu:2003dj}, including nonminimal coupling
to gravity, interactions between dark energy and the matter
sector, modified Lagrangians, braneworld models and quantum
effects. Many of these mechanisms require complicated additional
features and fine-tunings -- it is not easy to construct simple
and natural phantom-like behaviour without a phantom field. Among
the less complicated models is that of Csaki et
al.~\cite{Csaki:2004ha}, in which photon-to-axion conversion
mimics super-acceleration (in the presence of non-phantom dark
energy).

Phantom behaviour may also occur in braneworld models with an
infrared modification of general relativity, as pointed out by
Sahni and Shtanov~\cite{Sahni:2002dx}. A special case of the
Sahni-Shtanov models was further investigated by Lue and
Starkman~\cite{Lue:2004za}. In these models, the 4-dimensional
brane universe contains only matter and a cosmological constant
$\Lambda$, but a 5-dimensional gravitational effect leads to
phantom behaviour. The most important implication of this modified
gravity effect is that $\Lambda$ is effectively screened, so that
in principle a higher value of $\Lambda$ is allowed by the
observations than in the LCDM case. These models are the simplest
in the Shani-Shtanov class, with only one parameter more than the
LCDM model, and they are the simplest braneworld models with
phantom behaviour as far as we are aware. In this paper, we extend
the investigation of Lue and Starkman by testing the models
against observations, to see how the data constrains the model
parameters, and to compare the goodness of fit to that of LCDM. As
explained below, this test can only be partially carried out,
since the CMB anisotropies and matter power spectrum for the model
have not yet been computed, and this remains a formidable open
problem. However we can apply geometric observational tests that
are based on the background dynamics.

\section{The phantom-like braneworld}

The Dvali-Gabadadze-Porrati (DGP) braneworld model, generalized to
cosmology by Deffayet~\cite{Dvali:2000rv}, is a self-accelerating
model without any form of dark energy. In fact, the
self-accelerating models are not the only form of DGP model. There
are two separate branches, DGP$(\pm)$, depending on how the 4D
brane universe is embedded in the 5D spacetime.

\begin{itemize}
\item
The (+) branch is the self-accelerating model, which has recently
been tested against supernova (SN) and baryon oscillation (BO)
observations~\cite{Fairbairn:2005ue}, and then in addition against
the CMB shift data~\cite{Maartens:2006yt}. The flat DGP(+) model
is outside the 2$\sigma$ contour for the joint SN and BO
constraints, but this is somewhat misleading, since the CMB
observations have not been applied. When the CMB shift constraint
is included, the flat DGP(+) model is within the 2$\sigma$ contour
for the joint constraints, but the best-fit LCDM model gives a
significantly better fit to the data~\cite{Maartens:2006yt}.

\item
The DGP$(-)$ model is very different. It does not self-accelerate,
but requires dark energy on the brane. The simplest model has a
cosmological constant, and we will call this the LDGP model,
following Ref.~\cite{Lue:2004za}. The LDGP model is a special case
of the Sahni-Shtanov models with zero brane tension. It
experiences 5D gravitational modifications to its dynamics, which
effectively screen the cosmological constant. At late times, as
gravity leaks off the 4D brane, the dynamics deviates from general
relativity. The transition from 4D to 5D behaviour is governed by
a crossover scale $r_c$, as in the (+) branch. (The LDGP model has
recently been generalized by replacing the cosmological constant
with a quintessence field~\cite{Chimento:2006ac}.)

\end{itemize}

The energy conservation equation for LDGP remains the same as in
general relativity, but the Friedman equation is modified:
\begin{eqnarray}
&& \dot\rho+3H(\rho +p)=0\,,\label{ec} \\ && H^2+{H \over r_c}=
{8\pi G \over 3}\rho +{\Lambda \over 3}\,. \label{f}
\end{eqnarray}
[The self-accelerating (+) branch has $-H/r_c$ instead of
$+H/r_c$.] For the CDM case, with $p=0$, these equations imply
 \be
\dot H=-4\pi G\rho\left[ 1 - {1\over \sqrt{1+ 32\pi G
r_c^2\rho/3+4r_c^2\Lambda/3}} \right]. \label{r}
 \ee
Equation~(\ref{f}) shows that at early times, the general
relativistic Friedman equation is recovered:
 \be
H \gg r_c^{-1}~\Rightarrow~ H^2 \approx {8\pi G \over 3}\rho
+{\Lambda \over 3}\,.
 \ee
By contrast, at late times, the $H/r_c$ term is important and the
Friedman equation is nonstandard.

\begin{figure*}
\begin{center}
\includegraphics[width=7cm]{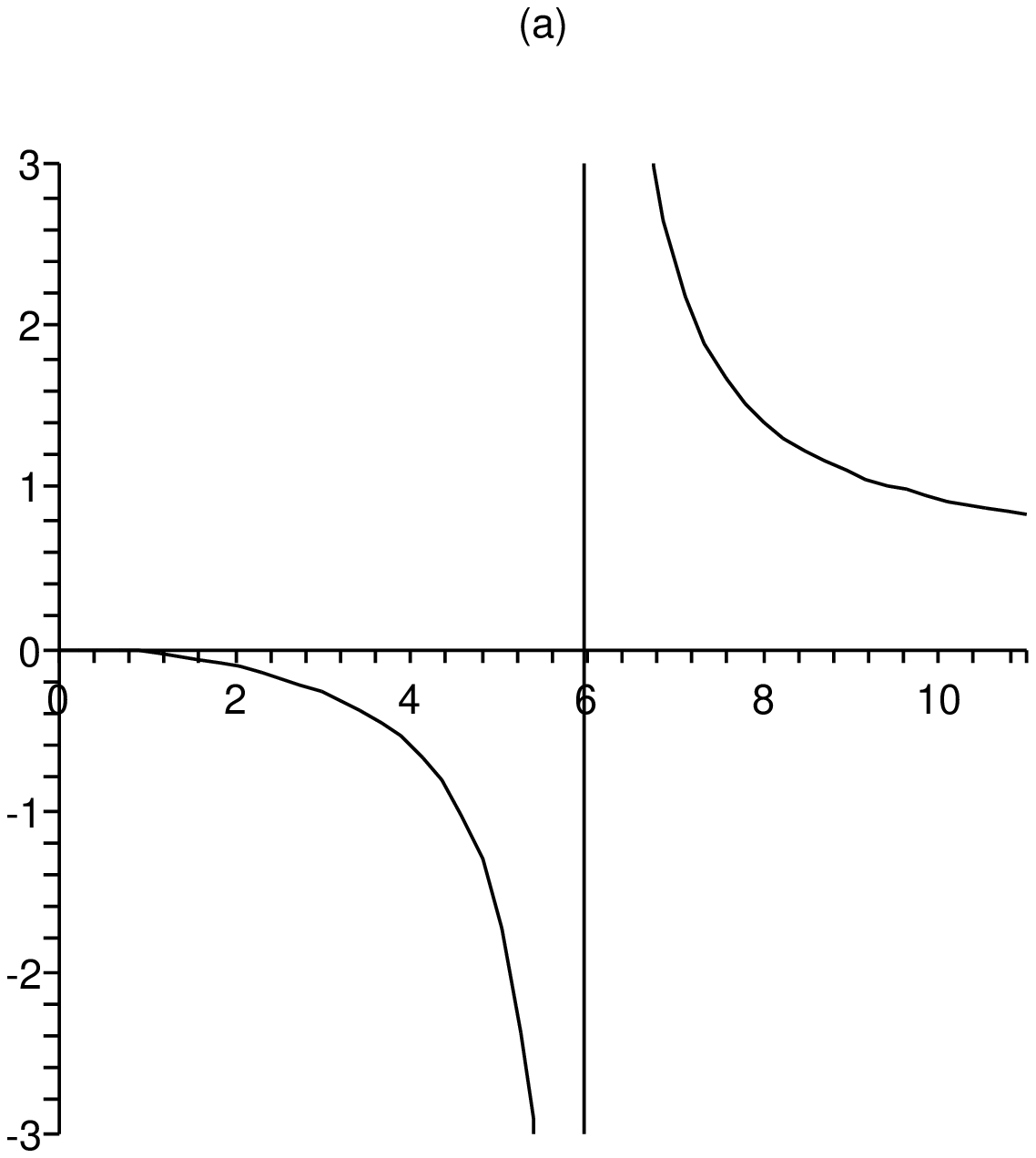}\quad
\includegraphics[width=7cm]{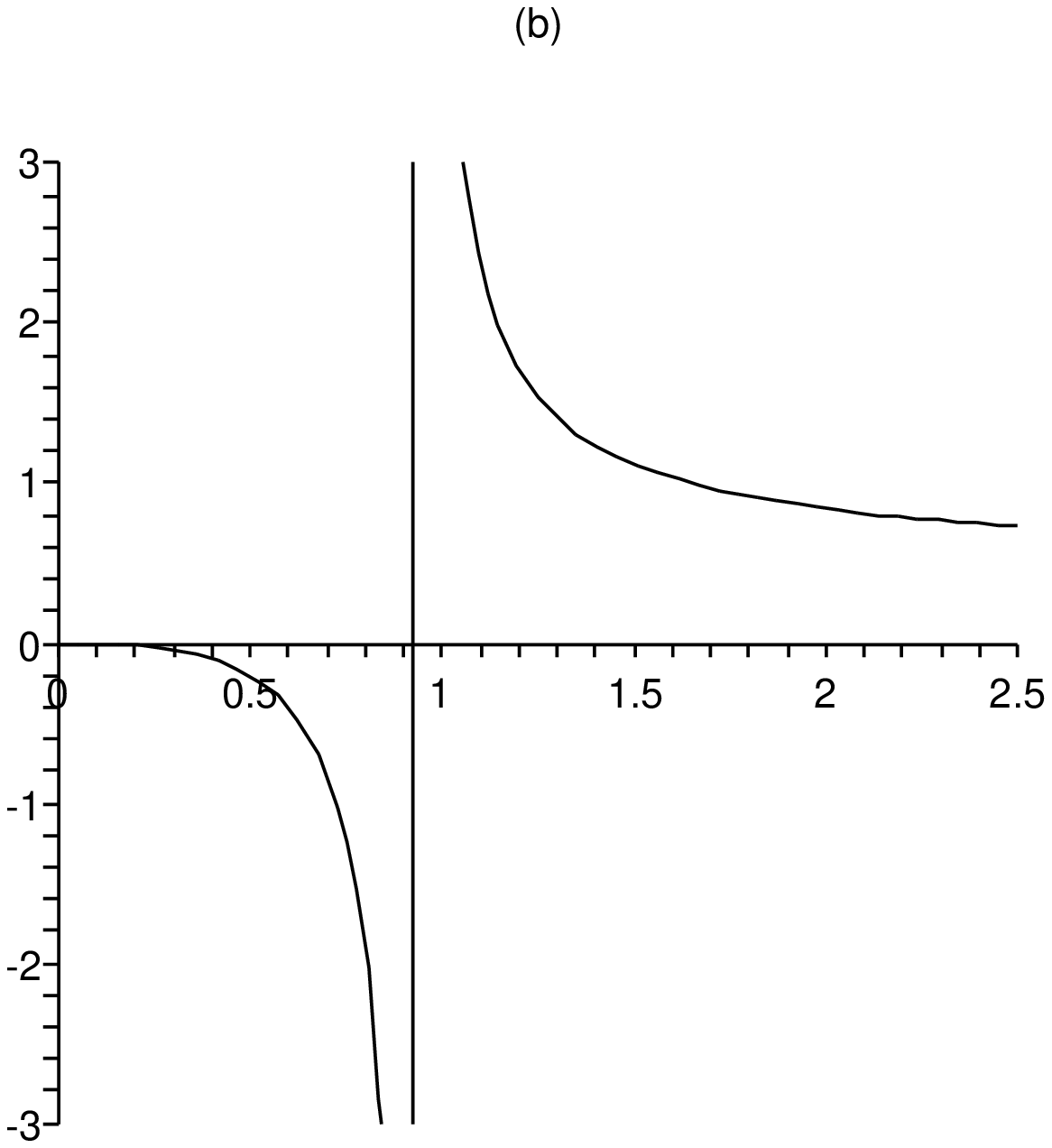}\quad \\
\includegraphics[width=7cm]{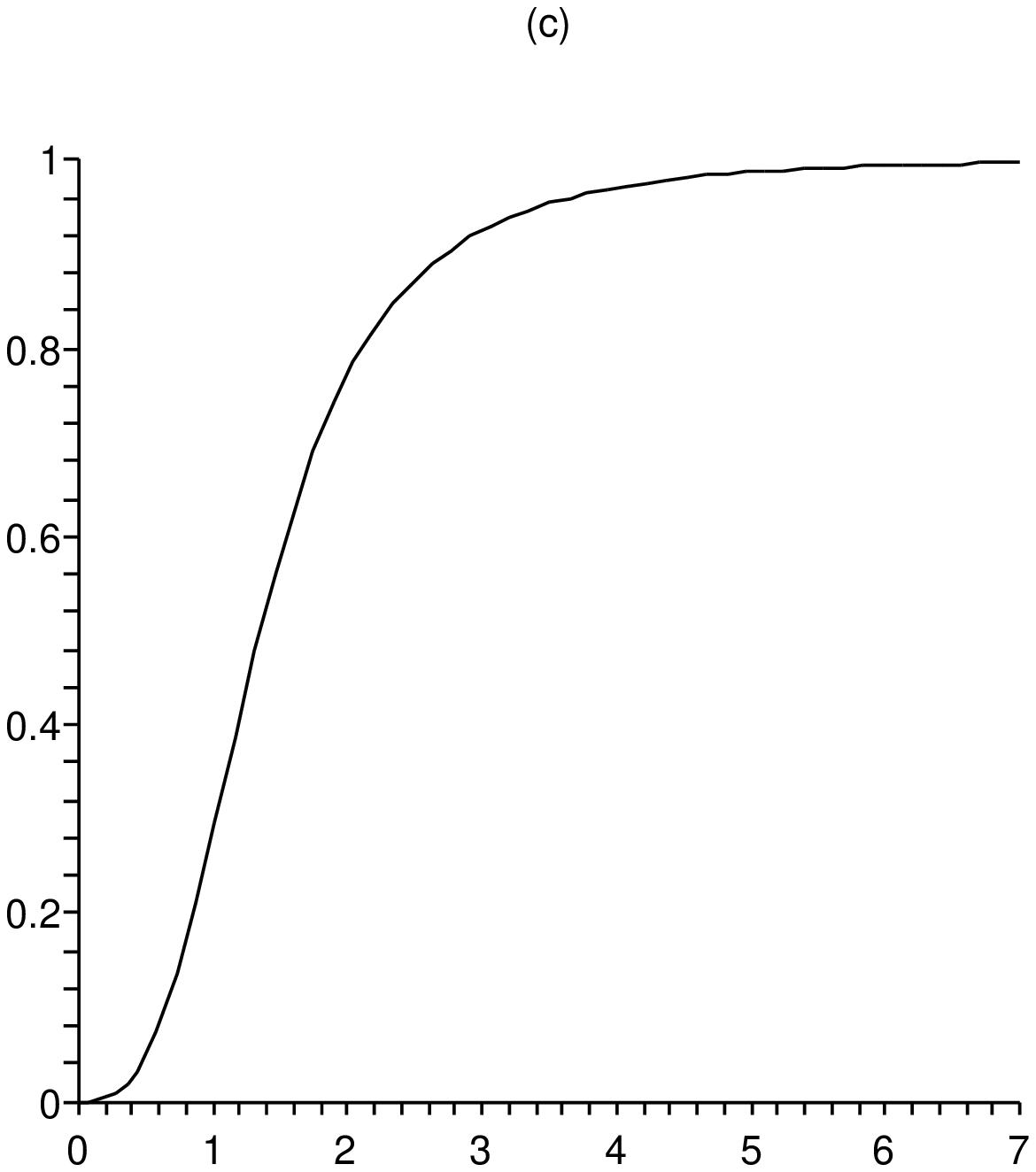}
\caption{(a) and (b): $1+w_{\text{eff}}$ against $1+z$. (c):
$1+w_{\text{tot}}$ against $1+z$, with the same parameters as
(a).\\ (a) shows a typical model with $w_{\text{eff}}<-1$ at
$z=0$, with parameters $(\om,\ol)=(0.3,0.8)$. (b) shows a model
with unrealistic parameters, $(\om,\ol)=(1.1,1.4)$, that has
$w_{\text{eff}}>-1$ at $z=0$. Note that $w_{\text{eff}}\to -1$ as
$z\to-1$ ($a\to\infty$), and $w_{\text{eff}}\to-{1\over 2}$ as
$z\to\infty$. The redshift $z_*$ where $w_{\text{eff}}$ goes
through $\pm\infty$ is clearly evident, with $z_*>0$ in (a) and
$z_*<0$ in (b). Note that $w_{\text{tot}}$ is continuous at $z_*$.
}\label{weff}
\end{center}
\end{figure*}

Gravity leakage at late times screens the cosmological constant,
leading to an effective dark energy~\cite{Sahni:2002dx,Lue:2004za}
 \be \label{rx}
\rx={1\over 8\pi G}\left(\Lambda-3{H \over r_c} \right),
 \ee
where
 \be
H^2= {8\pi G \over 3}(\rho+\rx).
 \ee
By Eq.~(\ref{r}), $\dot H<0$, so that
 \be
\dot{\rho}_{\text{eff}}>0\,.
 \ee
This is the basis for the phantom-like behaviour in the model. We
can define the effective dark energy equation of state
$w_{\text{eff}}= p_{\text{eff}}/\rx$ via
 \be \label{rp}
\dot{\rho}_{\text{eff}}+3H(1+w_{\text{eff}})\rx=0\,,
 \ee
or equivalently,
 \be \label{dh}
\dot H=-4\pi G [\rho+(1+w_{\text{eff}})\rx].
 \ee

Thus $\rx$ and $w_{\text{eff}}$, as defined in
Eqs.~(\ref{rx})--(\ref{dh}), are effective quantities that give a
standard general relativistic interpretation of LDGP expansion
history, i.e., they describe the equivalent general relativity
model. This equivalent model is a phantom model: by
Eqs.~(\ref{rx}) and (\ref{rp}), we have
 \be \label{wx}
1+w_{\text{eff}}={\dot H \over 8\pi G r_cH\rx}\,.
 \ee
Since $\dot H< 0$ by Eq.~(\ref{dh}), we have effective phantom
behaviour,
 \be\label{wx2}
w_{\text{eff}}<-1~~~(\rx>0)\,.
 \ee
The effective picture breaks down when $\rx=0$, i.e., when
$H=H_*=r_c\Lambda/3$, which always happens at some redshift $z_*$
in the history of the universe. At this time,
$w_{\text{eff}}\to-\infty$. At higher redshifts,
$1+w_{\text{eff}}$ is positive, decreasing from $+\infty$, as
illustrated in Fig.~\ref{weff}. This is a special case of the
behaviour found by Sahni and Shtanov~\cite{Sahni:2002dx}. Nothing
unphysical occurs at $z_*$, it is only the description that breaks
down. Note that it is possible that $z_*<0$, i.e., in the future,
so that $H_0>H_*$. But in this case, $w_{\text{eff}}$ at $z=0$ is
positive, so that there is no phantom behaviour for nonnegative
redshifts. In fact, Eq.~(\ref{wx0}) below shows that this only
occurs for the unphysical parameter values, $\om>1$.

Although $w_{\text{eff}}\to-{1\over 2}$ for large redshifts, CDM
is becoming strongly dominant, so that the geometrical constraints
from observations can be comfortably satisfied -- as we show in
the next section. As discussed further in the concluding section,
it remains an open question whether structure formation
constraints can also be satisfied.

The effective phantom behaviour in LDGP has no associated
instability, unlike phantom scalar fields in general relativity.
Furthermore, in general relativistic phantom models, $\dot H$
eventually becomes positive, i.e., the universe eventually
super-accelerates, which can lead to a ``big rip"
singularity~\cite{Nojiri:2005sx}. This also happens in a number of
models that mimic phantom behaviour without phantom matter. By
contrast, in LDGP, $\dot H$ is always negative, and there is no
big rip singularity in LDGP. Equation~(\ref{dh}) shows that the
key issue is the sign of $1+w_{\text{eff}}$ as $a\to\infty$. In
phantom and most phantom-like models, $1+w_{\text{eff}}<0$ as
$a\to\infty$, so that $\dot H$ becomes positive. In LDGP, we have
$1+w_{\text{eff}}\to 0^{-}$ and the universe is asymptotically de
Sitter:
 \be
w_{\text{eff}}\to -1\,, ~ H\to H_{\text{dS}}={1\over 2 r_c}\left[
\sqrt{{4 r_c^2 \Lambda \over 3}+1 }-1 \right].
 \ee
Note that $H_{\text{dS}}$ is less than the corresponding LCDM
value $H_{\text{dS,LCDM}}=\sqrt{\Lambda/3}$. This is an aspect of
the effective screening of $\Lambda$ by DGP($-$) gravity.

In LDGP, the total equation of state parameter is always greater
than $-1$, i.e., the phantom effects never dominate. Although the
screened dark energy has phantom equation of state,
$w_{\text{eff}}<-1$, the total $w$ remains above $-1$, and there
is {\em no} phantom-like acceleration of the universe, and thus no
big rip. This is shown by Eq.~(\ref{wt}) below, and illustrated in
Fig.~\ref{weff}.

\section{Observational constraints on LDGP}

\begin{figure*}
\begin{center}
\includegraphics[width=5cm]{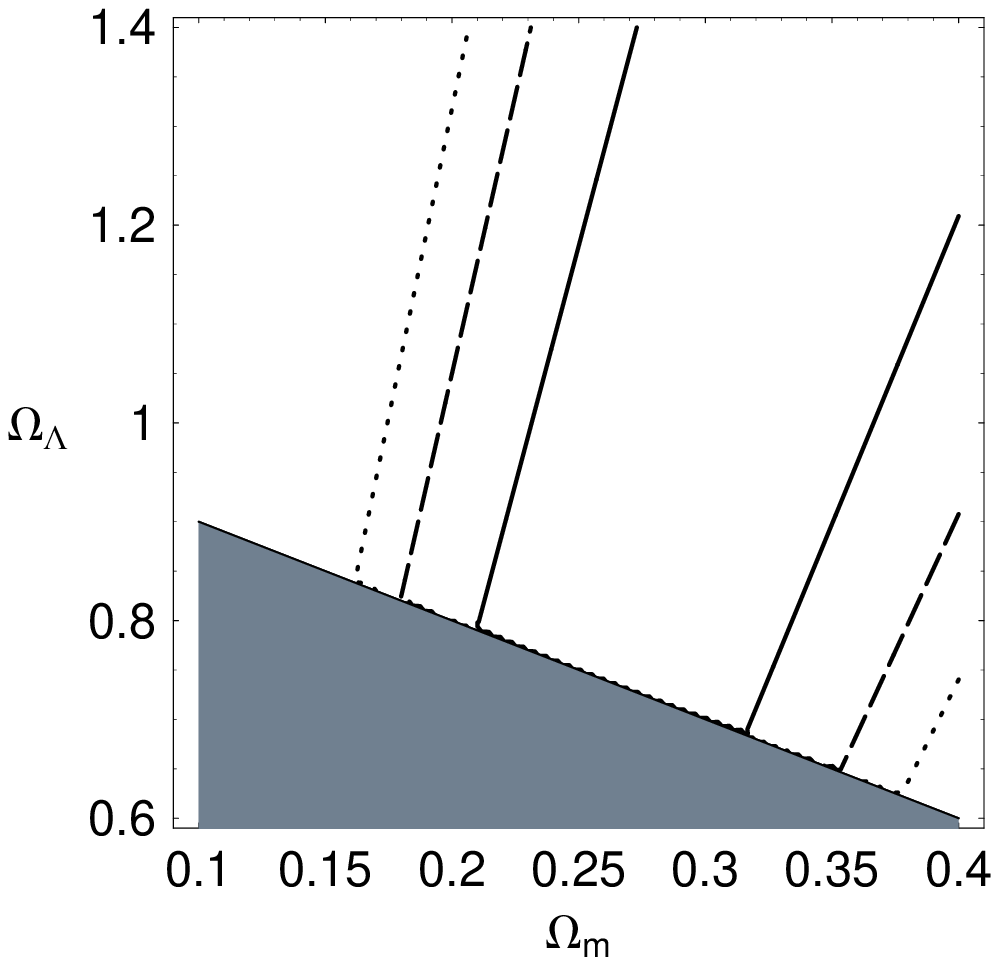}\quad
\includegraphics[width=5cm]{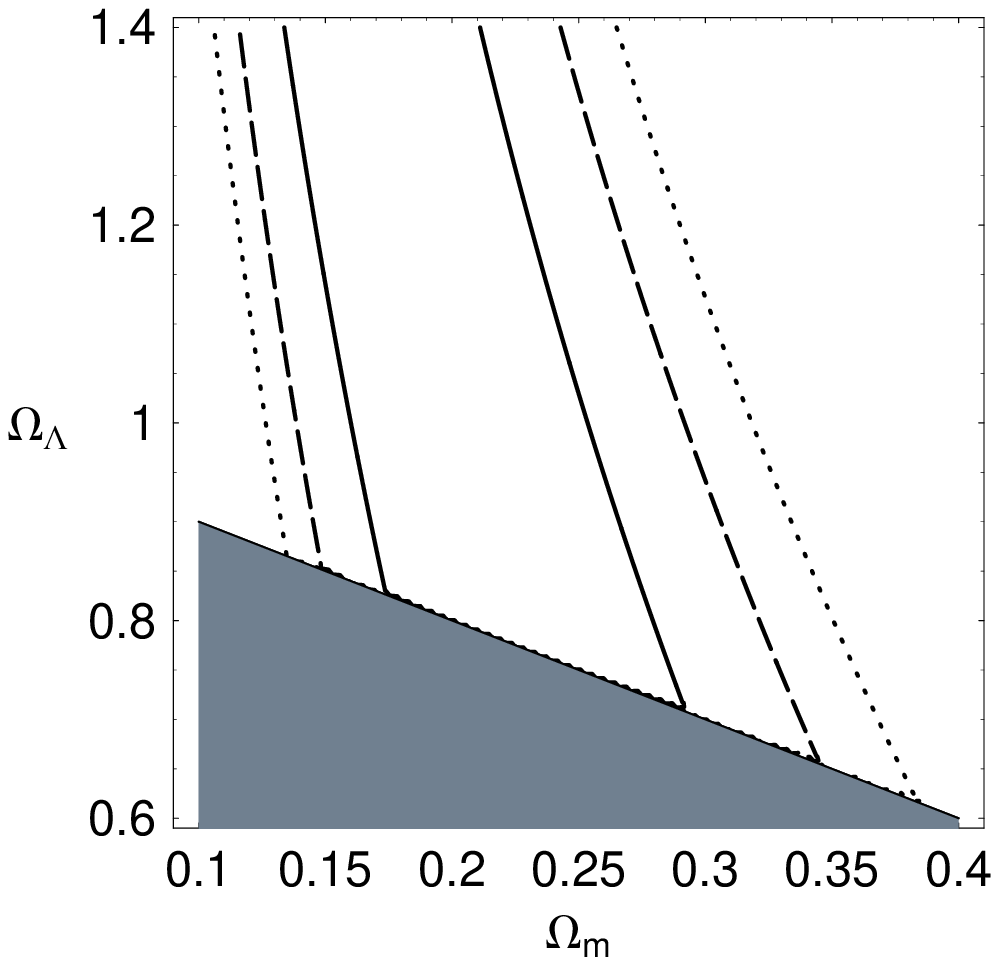}\quad
\includegraphics[width=5cm]{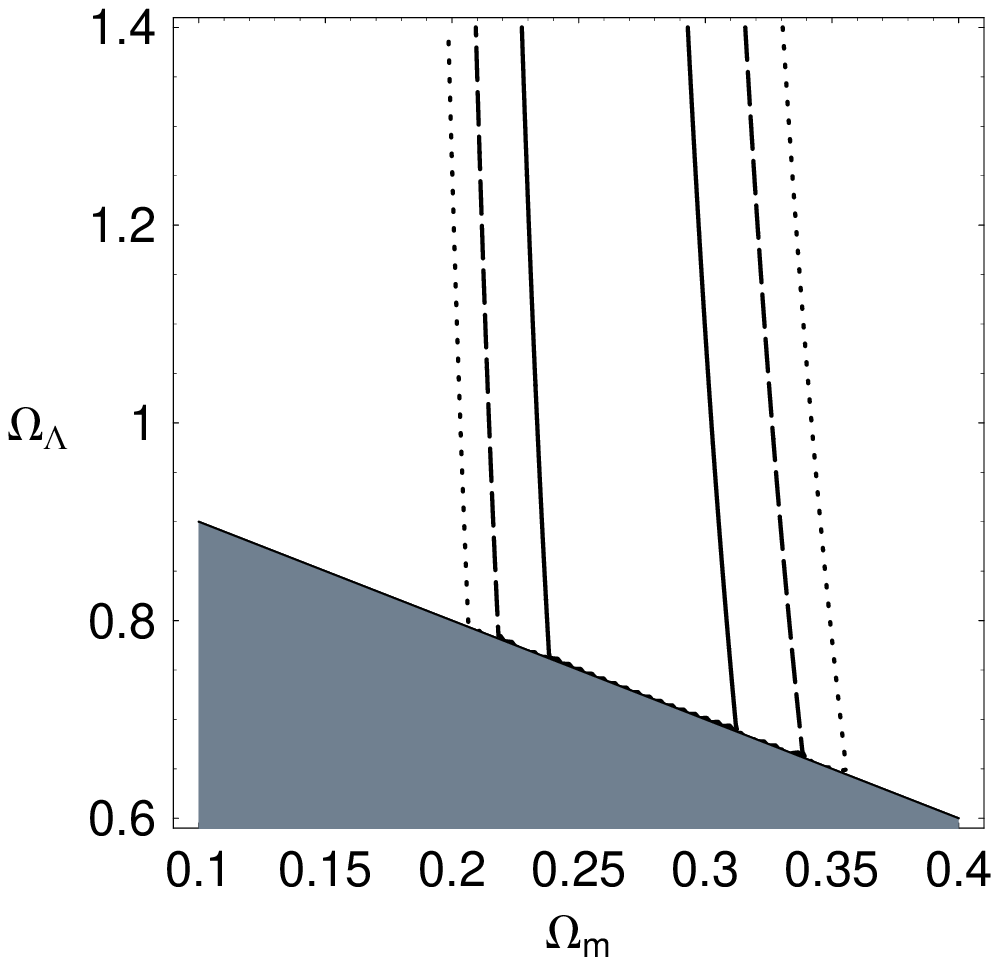}
\caption{ The 68\%, 95\% and 99\% confidence contours for the SNLS
data (left), the CMB shift parameter $S$ (based on WMAP 3-year
data) (middle) and the baryon oscillation measure $A$ (right). The
shaded area is the unphysical region, $\om+\ol<1$, and the
boundary is the flat LCDM model, $\orc=0$. }\label{shift}
\end{center}
\end{figure*}

In dimensionless form, the LDGP Friedman equation~(\ref{f})
implies
 \bea
E(z)\!\equiv\! {H(z) \over H_0}= \!\sqrt{\om (1+z)^3+
\ol+\orc}-\!\sqrt{\orc} , \label{fz}
 \eea
where
 \bea
\om &=& {8\pi G \rho_0 \over 3H_0^2}\,,~ \ol = {\Lambda \over
3H_0^2}\,,\\ \orc &=&{1 \over 4 H_0^2 r_c^2}\,.
 \eea
Note that Eq.~(\ref{f}) implies
 \be \label{omol}
\om+\ol=1+2\sqrt{\orc}\,,
 \ee
so that in particular,
 \be
\om+\ol \geq 1\,.
 \ee
This condition ensures that $H_0^2\geq 0$. The flat LCDM model is
the limiting case of equality, when $\orc=0$. The unphysical
region $\om+\ol<1$ (corresponding to imaginary $H_0$) is shown in
Fig.~\ref{shift}.

Equation~(\ref{omol}) means that the flat LDGP model has the
remarkable feature that it mimics a {\em closed} LCDM model in the
$(\om,\ol)$ plane:
 \be \label{orok}
\sqrt{\orc} \leftrightarrow -{\Omega_K \over 2}\,.
 \ee
The WMAP 3-year data with SN data gives~\cite{Spergel:2006hy}
 \be \label{okol}
\Omega_K=-0.01^{+.016}_{-.009}\,,~~ \ol=0.72 \pm .04\,.
 \ee
If we take the mean value, $\Omega_K=-0.01$, then Eq.~(\ref{orok})
gives the ``equivalent" value $r_c=100H_0^{-1}$. This very large
value of $r_c$ suggests that observations will require the LDGP
model to be close to the LCDM model ($r_c=\infty$). We confirm
this expectation below.

By Eqs.~(\ref{wx}) and (\ref{fz}), we have
 \be
1+w_{\text{eff}}(z)= -{\sqrt{\orc}\om(1+z)^3 \over
[\ol-2\sqrt{\orc}\,E(z)] [\sqrt{\orc}+E(z)]}\!.\label{wx3}
 \ee
This equation shows explicitly how we can mimic the expansion
history of LDGP by an effective phantom model in general
relativity. However, there is no violation of the null energy
condition, and no super-acceleration, since in LDGP the phantom
dynamics is gravitational. Figure~\ref{weff} illustrates this
equation.

At the current time,
 \be \label{wx0}
1+w_{\text{eff}}(0)=-{(\om+\ol-1)\om \over (1-\om)(\om+\ol+1)}.
 \ee
Provided that $\om<1$, we have phantom behaviour at the current
time, i.e., Eq.~(\ref{wx2}) is satisfied.

As noted above, the phantom effects never dominate, and the total
equation of state parameter, $w_{\text{tot}}=p_{\text{tot}}/
\rho_{\text{tot}}=w_{\text{eff}}\rx/(\rho+\rx)$, is always greater
than $-1$. By Eqs.~(\ref{fz}) and (\ref{wx3}),
 \bea
&& 1+w_{\text{tot}}(z)= {\om (1+z)^3 \over E(z)\left[
\sqrt{\orc}+E(z)\right]}\,,\label{wt}
 \eea
and it follows immediately that $w_{\text{tot}}(z)\geq -1$. Since
$E \to \sqrt{\ol+\orc}-\sqrt{\orc}$ as $z\to -1$ (i.e., as
$a\to\infty$), it follows that $1+ w_{\text{tot}} \to 0^+$. At
early times, i.e., $z\to\infty$, we have $w_{\text{tot}} \to 0$,
which is the GR limit (we are neglecting radiation). This
behaviour is illustrated in Fig.~\ref{weff}.

From Eq.~(\ref{r}), the dimensionless acceleration is
 \be
{\ddot a/a\over H_0^2}=E\!\left[\!{2\orc +2\ol -\om (1+z)^3 \over
2\sqrt{\om (1+z)^3+ \ol+\orc}}\! -\! \sqrt{\orc} \right],
 \ee
so that the redshift when acceleration starts is given by
 \bea
&&\!\!\!\!\!\!\! 1+z_{\text{acc}}=
\nonumber\\&&\!\!\!\!\left(2{\ol \over \om}\right)^{\!1/3}
\left[\!1+2{\orc\over \ol}\!\left(\!1-\!\sqrt{1+{3\ol \over 4\orc}
}\, \right)\right]^{\!1/3} \!.\label{za}
 \eea
The LCDM result is recovered for $\orc=0$. When
$z_{\text{acc}}=0$, this gives the critical line $\ol=2\om-1$, so
that
 \be
\om<{1\over 2}(1+\ol)\,,
 \ee
is the condition for models that are currently accelerating. This
differs from the corresponding condition for LCDM, i.e.,
$\om<2\ol$.

The fundamental test of the background dynamics of a cosmological
model is the SN magnitude-redshift test, based on the luminosity
distance,
 \bea
d_L &=& {(1+z) \over H_0}\, \int_0^z\, {dz' \over E(z')} \,.
\label{ld}
 \eea
(Here we are restricting to the flat case.) The 68\%, 95\% and
99\% confidence contours from fits to the Legacy (SNLS)
data~\cite{Astier:2005qq} are shown in the LDGP parameter plane in
Fig.~\ref{shift}.

Further independent tests are needed to check whether the LDGP
model is consistent with the observed features of the universe.

\begin{figure}
\begin{center}
\includegraphics[width=8cm]{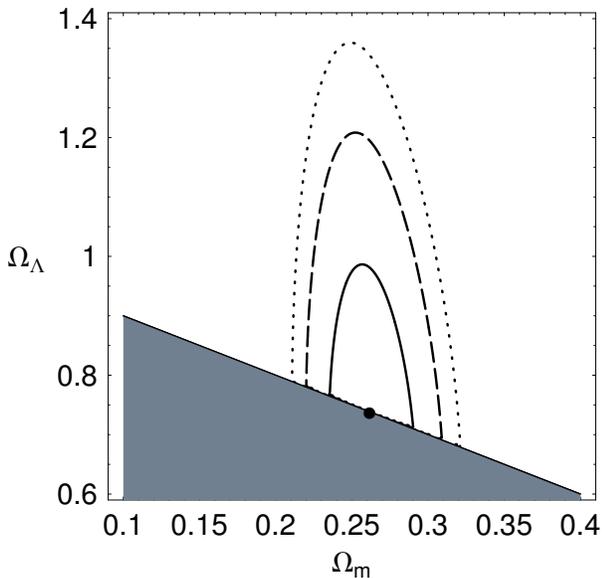}
\caption{Joint constraints on LDGP models from the SNLS data, the
BO measure $A$ and the CMB shift parameter $S$. The solid dot is
the best-fit model. The shaded area is the unphysical region,
$\om+\ol<1$, and the boundary is the flat LCDM model, $\orc=0$.
}\label{sac}
\end{center}
\end{figure}

The CMB shift parameter
 \bea
S = \sqrt{\om}H_0\,{d_L(z_{\text{r}}) \over (1+z_{\text{r}})} \,,
  \eea
encodes the relation between the angular diameter distance to last
scattering, the angular scale of the first acoustic peak, and the
physical scale of the sound horizon. This parameter is effectively
model-independent and provides a good test of the background
dynamics, independent of the SN redshift test. We take
$z_{\text{r}}=1090$. Wang and Mukherjee~\cite{Wang:2006ts} have
used the WMAP 3-year data to find that
 \be
S= 1.70~ \pm 0.03\,.
 \ee
This value shows a significant improvement in the error over the
value from the 1-year data~\cite{Wang:2003gz}, $S=1.72 \pm 0.06$.
The constraints from the CMB $S$ parameter, based on the 3-year
data, are shown in Fig.~\ref{shift}.

\begin{figure*}
\begin{center}
\includegraphics[width=7.5cm]{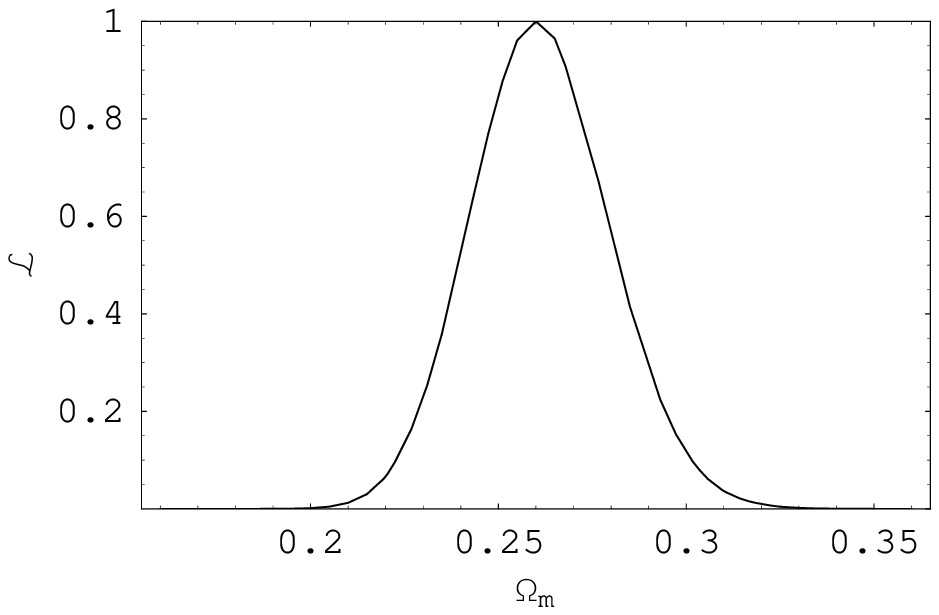}\quad\quad
\includegraphics[width=7.5cm]{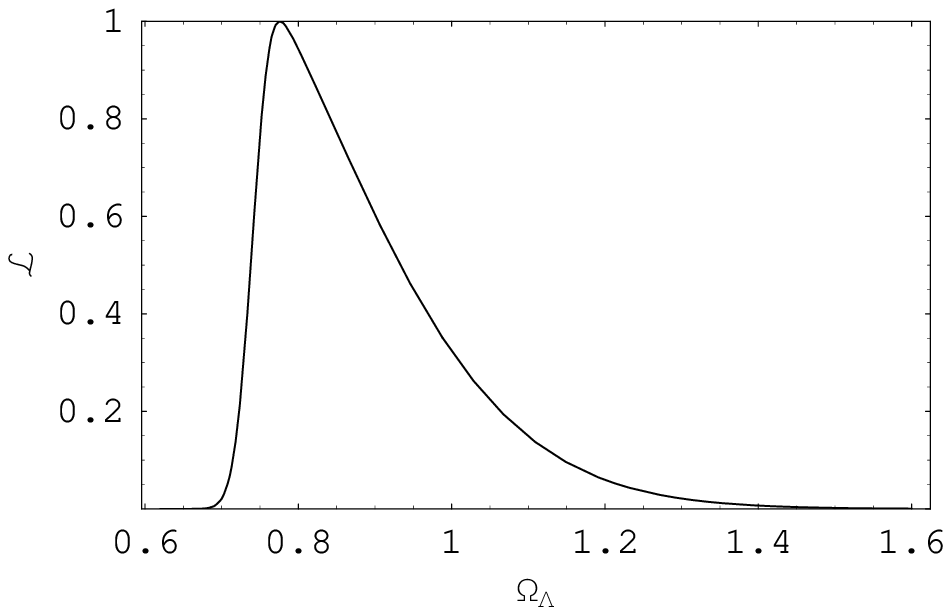}
\caption{The normalized marginalized likelihoods for $\om$ and
$\ol$, subject to the joint constraints of data from SNLS
supernova redshifts, CMB shift parameter $S$, and baryon acoustic
oscillation measure $A$. }\label{lik}
\end{center}
\end{figure*}

The baryon acoustic oscillation peak recently detected in the SDSS
luminous red galaxies (LRGs)~\cite{Eisenstein:2005su} can be used
as another independent geometrical test of the LDGP. The
correlation function for SDSS LRGs shows a peak at a scale $\sim
100 h^{-1}\,$Mpc, corresponding to the first acoustic peak at
recombination (determined by the sound horizon). The observed
scale effectively constrains the quantity~\cite{Eisenstein:2005su}
 \bea
A &=& \sqrt{\om}\left[ {H_0^3\, d_L^2(z_1) \over H_1
z_1^2(1+z_1)^2} \right]^{1/3},
 \eea
where $z_1=0.35$ is the typical LRG redshift. Eisenstein et
al.~\cite{Eisenstein:2005su} give
 \be
A = 0.469 \pm 0.017\,.\label{boa}
 \ee
(We have suppressed a weak dependence of $A$ on the spectral
tilt.) Note that there is a level of uncertainty in the use of the
BO measure $A$ to test non-LCDM models. This uncertainty will
require substantial further effort, both in the processing of the
data~\cite{Maartens:2006yt,Dick:2006ev} and in the theoretical
understanding of the LDGP matter power spectrum (see
Ref.~\cite{Maartens:2006yt} and the discussion below). For the
present, we use the constraint in Eq.~(\ref{boa}), and assume that
the unresolved issues will lead to only small corrections.
Constraints from the BO data are shown in Fig.~\ref{shift}.

The joint constraints from SN, CMB shift and BO data are shown in
Fig.~\ref{sac}.

\section{CONCLUSIONS}

A naive comparison between LDGP models and closed LCDM models, as
illustrated in Eqs.~(\ref{orok}) and (\ref{okol}), already
suggests that the data will favour LDGP models that are close to
the flat LCDM model. The joint constraint contours in
Fig.~\ref{sac} confirm this expectation.

The best fit from the joint constraints is
 \be
\om=0.26\,,~\ol=0.74\,,~ w_{\text{eff}}(0)=-1.00\,,
 \ee
with $\chi^2$ of
 \be
\chi^2=114.6\,,~~\chi^2~\text{per degree of freedom}\,=0.996\,.
 \ee
Interestingly, the best-fit LDGP model, with current data and
within the accuracy of our calculations, is the LCDM limit of
LDGP. Since the LCDM model has one less parameter, it provides a
better fit as measured by the reduced $\chi^2$:
 \be
\chi^2\Big|_{\text{LCDM}}~\text{per degree of freedom}\,=0.988\,.
 \ee
It is the CMB shift value from the 3-year WMAP data that is
forcing the best-fit strongly towards the LCDM limit: if we use
the shift value from the 1-year data~\cite{Wang:2003gz}, i.e.,
$S=1.72 \pm 0.06$, then the best-fit model from the joint
constraints moves off the LCDM boundary:
 \be
\om=0.27\,,~\ol=0.78\,,~w_{\text{eff}}(0)=-1.01~(\text{1-year
}\,S)\,,
 \ee
with $\chi^2=113.7$, and a reduced $\chi^2$ per degree of freedom
of $0.988$.

The best-fit LDGP model happens to be an LCDM model, using the
current data, but it is more meaningful to give the 68\%
confidence limits on the mean values of the parameters. The 68\%
contour in Fig.~\ref{sac} shows that a broad range of non-LCDM
models is consistent with the joint data constraints. In
principle, we can calculate this range by marginalizing in turn
over $\ol$ and $\om$. In practice this is made difficult by the
asymmetrical shape of the physical region in parameter space. The
marginalized likelihoods are shown in Fig.~\ref{lik}. The $\om$
likelihood has a near-gaussian shape, so that the median gives a
good measure of the mean. Then we can calculate the limits by
integrating to 34.1\% of the area to the left and 34.1\% to the
right. This leads to
 \be\label{lik1}
\om=0.261^{+0.019}_{-0.018}\,.
 \ee
By contrast, the $\ol$ likelihood curve is strongly asymmetrical,
reflecting the fact that the unphysical region has a stronger
effect on $\ol$ than on $\om$. As a consequence, the mode and the
median are significantly different, and we cannot extract a
meaningful mean value for $\ol$. A conservative upper limit on
$\ol$ may be computed by integrating from the mode until 68\% of
the area under the curve. This gives
 \be\label{lik2}
\ol < 0.95~\text{ at 68.3\% confidence}\,.
 \ee

\begin{figure*}
\begin{center}
\includegraphics[width=7.5cm]{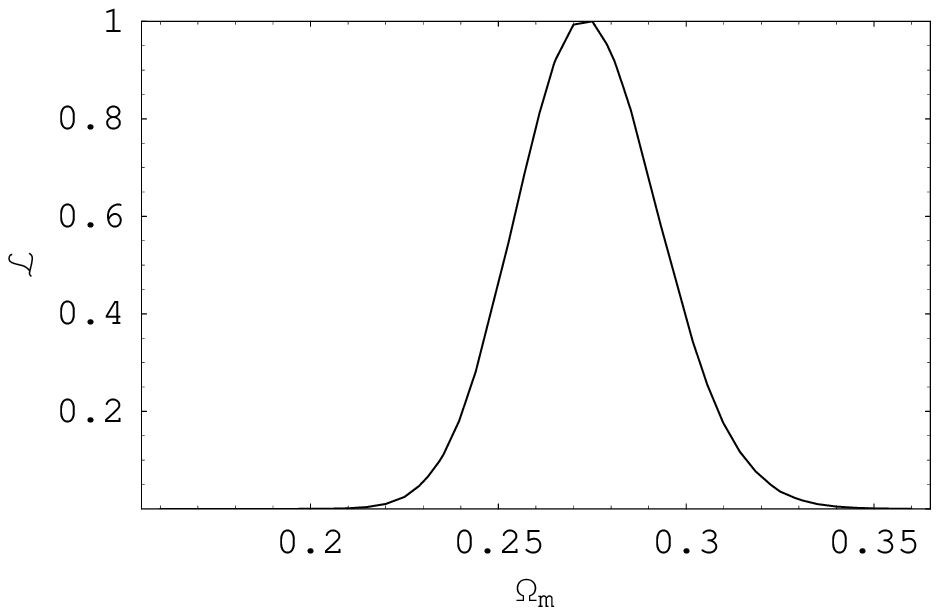}\quad\quad
\includegraphics[width=7.5cm]{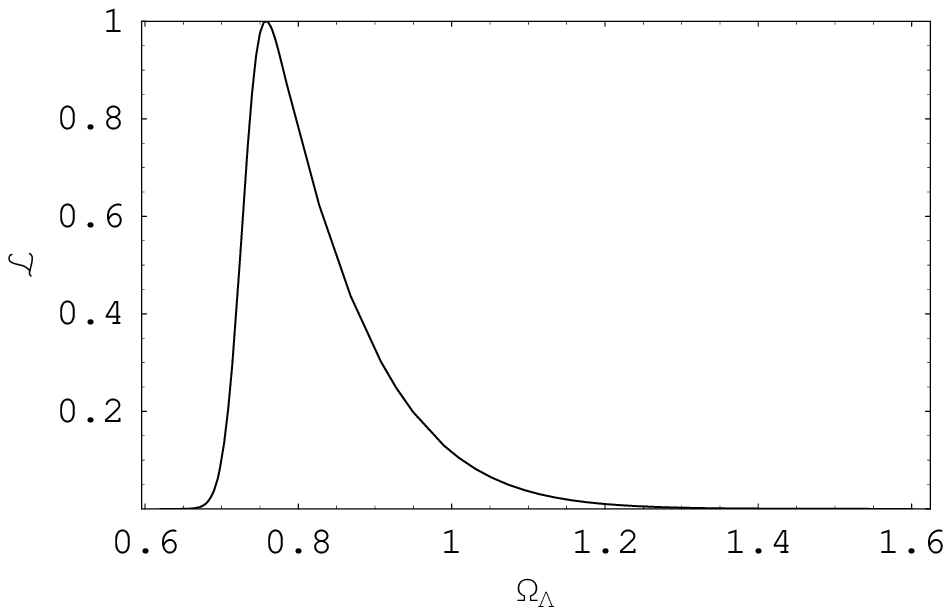}
\caption{As in Fig.~\ref{lik}, but using the Gold supernova data.
}\label{likg}
\end{center}
\end{figure*}

If we use the Gold SN data~\cite{Riess:2004nr} instead of the SNLS
data, then there are some interesting changes in the values of the
parameters, but the best fit lies still on the LCDM line. The Gold
data favours a higher best-fit $\om$:
 \bea
\!\!\!\!\!&& \om=0.275\,,~ \ol=0.725\,,\\\!\!\!\!\! &&
\chi^2=179.3\,,~ \chi^2~\text{per degree of freedom}\, = 1.142\,.
 \eea
For LCDM the $\chi^2$ per degree of freedom is 1.135. The
marginalized likelihoods for $\om$ and $\ol$ are shown in
Fig.~\ref{likg}. We find that
 \bea
&& \om=0.274^{+0.019}_{ -0.019}\,, \label{likg1}\\ &&
\ol<0.87~\text{ at 68.3\% confidence}\,.\label{likg2}
 \eea
Thus the Gold data gives a stronger upper limit on $\ol$.

Equations~(\ref{lik1}), (\ref{lik2}) and (\ref{likg1}),
(\ref{likg2}) summarize our quantitative conclusions about the
observational constraints on LDGP models. The 3 independent data
sets allow for a significant range of these phantom-like
braneworld models, with $\om$ values that are consistent with
other observational tests, and with $\ol$ values that imply
significant screening of $\Lambda$. However, the LCDM model, with
one less parameter, does provide a better fit to the data. The
evidence for braneworld screening of $\Lambda$, via effective
phantom behaviour, is therefore weak. This is consistent with
results for general relativistic models. If one uses only
supernova observations, especially the Gold
data~\cite{Riess:2004nr}, then there appears to be evidence for
phantom dark energy. But it is misleading to ignore the CMB
observations, and when these are imposed, the evidence for phantom
behaviour is much weaker~\cite{Jassal:2006gf}.

It is important to stress that we have only imposed observational
tests on the background dynamics of LDGP. In regard to background
dynamics, which are determined by $H(z)$, the LDGP model is
indistinguishable from a GR model with dark energy whose equation
of state is exactly $w_{\text{eff}}(z)$, as given by
Eq.~(\ref{wx3}). However, the GR model has no physical motivation,
whereas the LDGP model gives a covariant and consistent physical
meaning to the equation of state~(\ref{wx3}). Nevertheless, one
needs to break this degeneracy. The way to do this is to impose
observational tests based on structure formation.

Why have we not done this, i.e., computed the detailed CMB
anisotropies and matter power spectrum for the LDGP model? The
reason is that the density perturbation equations for LDGP are
5-dimensional, and their solution is still a formidable unsolved
problem. An approximate solution to the 5D equations in the DGP(+)
models has been developed for small, sub-Hubble scales, by Koyama
and Maartens~\cite{Koyama:2005kd}, confirming the intuitively
motivated solution of Lue and Starkman~\cite{Lue:2005ya}. This
approximate solution may also be applied to the LDGP case via the
replacement $r_c \to -r_c$, as pointed out by Lue and
Starkman~\cite{Lue:2004za}. But perturbations on scales near to
and above the Hubble scale feel increasingly strong 5D gravity
effects, and the approximation in
Refs.~\cite{Koyama:2005kd,Lue:2005ya} breaks down. The
perturbation problem becomes strongly 5D and the Fourier modes
satisfy partial rather ordinary differential equations. In the
DGP(+) case, the CMB anisotropies and matter power spectrum have
been presented in Refs.~\cite{wrong}. These papers describe
valuable strategies for distinguishing DGP from GR dark energy.
However, as explained by Koyama and Maartens~\cite{Koyama:2005kd},
their quantitative results are unreliable, since they ignore 5D
effects on density perturbations (which turns out to violate the
4D Bianchi identity).

Finally, we comment on the DGP ghost issue. In self-accelerating
DGP(+) models, the asymptotic de Sitter state of the model suffers
from a ghost~\cite{Gorbunov:2005zk}. However, in the LDGP case,
which is based on the DGP$(-)$ branch, there is no
ghost~\cite{Charmousis:2006pn}.

\[ \]
{\bf Acknowledgements:} RL is supported by the Spanish Ministry of
Science and Education through the RyC program, and research grants
FIS2004-01626 and FIS2005-01181. The work of RM is supported by
PPARC. We thank Rob Crittenden, Kazuya Koyama, Pia Mukherjee,
Israel Quiros, Varun Sahni, Yun Wang and Jussi Valiviita for
useful discussions.

\end{document}